\documentclass[11pt,a4paper]{article}
\usepackage{jcappub}

\usepackage{hyperref}
\usepackage{braket}
\usepackage{mathtools}
\usepackage{amsmath}
\usepackage{amsfonts}
\usepackage[dvipsnames]{xcolor}
\usepackage{graphicx}
\usepackage{caption}
\usepackage{float,subcaption}
\usepackage{amssymb}
\usepackage{tikz}
\usepackage{stackengine}
\usepackage{array}
\usepackage{booktabs,multirow}

\begin{document}

\title{Misalignment mechanism for a mass-varying vector boson}
\author[a]{Kunio Kaneta,}
\emailAdd{kkaneta@lab.twcu.ac.jp}
\author[b]{Hye-Sung Lee,}
\emailAdd{hyesung.lee@kaist.ac.kr}
\author[b]{Jiheon Lee,}
\emailAdd{anffl0101@kaist.ac.kr}
\author[b]{Jaeok Yi}
\emailAdd{wodhr1541@kaist.ac.kr}
\affiliation[a]{Department of Mathematics, Tokyo Woman’s Christian University\\ Tokyo 167-8585, Japan}
\affiliation[b]{Department of Physics, Korea Advanced Institute of Science and Technology\\ Daejeon 34141, Korea}
\abstract{
A coherent field over the entire universe is an attractive picture in studying the dark sector of the universe. The misalignment mechanism, which relies on inflation to achieve homogeneousness of the field, is a popular mechanism for producing such a coherent dark matter.
Nevertheless, unlike a scalar field case, a vector boson field suffers because its energy density is exponentially suppressed by the scale factor during the cosmic expansion.
We show that if the vector field gets a mass from a scalar field, whose value increases by orders of magnitude, the suppression can be compensated, and the misalignment can produce the coherent vector boson that has a sizable amount of energy density in the present universe. Quintessence can be such a scalar field.
}
\maketitle
\flushbottom

\section{Introduction} 
\label{Sec:intro}

The $\Lambda$-CDM model successfully explains many cosmological observational data, including the cosmological microwave background (CMB) spectrum \cite{Planck:2018vyg}. It consists of ordinary matter, cold dark matter (CDM), and dark energy described by the cosmological constant $\Lambda$. Because of its success, it is sometimes called the concordance cosmological model. Nonetheless, the $\Lambda$-CDM model bears many questions about its dark components. There have been copious studies about dark matter \cite{Silveira:1985rk,McDonald:1993ex,Burgess:2000yq,Djouadi:2011aa,Djouadi:2012zc,Mambrini:2011ik,Alves:2013tqa,Lebedev:2014bba,Arcadi:2013qia,Arcadi:2014lta,Ellis:2017ndg,Escudero:2016gzx,Hall:2009bx,Bhattacharyya:2018evo,Kaneta:2016vkq,Kaneta:2016wvf,Anastasopoulos:2020gbu,Brax:2020gqg,Brax:2021gpe,Kaneta:2021pyx,Chowdhury:2018tzw,Kamada:2018zxi}, but dark energy remains the vaguest part of the model.

In the $\Lambda$-CDM model, the $\Lambda$ is determined to fit the observational data including the CMB anisotropy \cite{Planck:2018vyg}. Although this description is simple, it is not satisfying for several reasons. For example, there is no particular reason why the energy density of the cosmological constant is comparable to that of dark matter in the recent era (cosmological coincidence problem) \cite{Velten_2014}. Also, the distant conjecture on the swampland prefers a dynamically varying vacuum to a constant vacuum \cite{Ooguri:2018wrx,Garg:2018reu}. In recent years, the Hubble tension, a $5\sigma$ discrepancy between the early universe measurement by Planck \cite{Planck2020} and the late universe measurement by SH0ES \cite{Riess:2021jrx} on the Hubble constant, has emerged as a significant challenge to the $\Lambda$-CDM model \cite{Verde:2019ivm,DiValentino:2021izs,Perivolaropoulos:2021jda,DiValentino:2022fjm,Schoneberg:2021qvd}. Since it is a deviation between the experimental data and $\Lambda$-CDM model, this tension triggered a sharp increase in dark energy research \cite{DiValentino:2017iww,Yang:2018euj,Poulin:2018cxd,Vagnozzi:2019ezj,DiValentino:2019ffd,DiValentino:2019jae,Jedamzik:2020zmd,Choi:2020pyy,Dainotti:2021pqg,Vagnozzi:2021gjh,Mawas:2021tdx,Dainotti:2022bzg,Yao:2023jau,Zhai:2023yny}. Therefore, exploring alternatives like quintessence and other dark energy scenarios is warranted.

Quintessence is a dark energy model identifying dark energy as a scalar field slowly rolling down the potential \cite{Wetterich:1987fm,PhysRevD.37.3406,Caldwell:1997ii}. Its dynamics may address the cosmological coincidence problem \cite{Zlatev:1998tr}. Also, the quintessence model can be modified to have some symmetries. For instance, the quintessence is formulated as the gauge field \cite{Pilo:2003gu,Mehrabi:2015lfa,Gevorkyan:2018zoh,Hill:2002kq}, associated with the global symmetry \cite{Gu:2001tr,Boyle:2001du,Li:2001xaa,Mainini:2004he,Frieman:1995pm,Kim:1998kx,Choi:1999xn,Kim:2002tq,Carroll:1998zi}, and Higgs field with a non-abelian gauge symmetry \cite{Rinaldi:2014yta,Rinaldi:2015iza,Alvarez:2019ues,Orjuela-Quintana:2020klr}. Although the idea of quintessence is very interesting, the minimal quintessence model usually worsens the Hubble tension  \cite{DiValentino:2017iww,DiValentino:2016hlg,Joudaki:2017zhq,Colgain:2019joh,Mawas:2021tdx,Heisenberg:2022lob,Heisenberg:2022gqk,Lee:2022cyh,Yao:2023jau,Zhai:2023yny} while a coupled quintessence might give a different result \cite{Banerjee:2020xcn,Lee:2022cyh}.

We suggested the ``gauged quintessence" model in the previous work \cite{Kaneta_2023}. This model interprets the quintessence as the radial part of a complex scalar charged by a dark Abelian gauge symmetry. The mass of a dark gauge boson, the vector gauge boson corresponding to the dark $U(1)$ symmetry, is given by the value of quintessence. Since the value of quintessence varies in time, the dark gauge boson has a mass-varying characteristic. Unlike the uncoupled quintessence model, the effective equation of state of the dark energy in the gauged quintessence model can be smaller than $-1$ in the near-past. Due to this fact, the gauged quintessence might relax the drawbacks of quintessence on Hubble tension. Also, recent studies suggest that scenarios containing the couplings between dark energy and dark matter could be a viable solution to Hubble tension \cite{DiValentino:2017iww,Yang:2018euj,DiValentino:2019ffd,Zhai:2023yny,Yao:2022kub}.

Therefore, it is important to ask ``How much relic dark gauge boson could be in the present?'' but the answer to this question is non-trivial due to the mass varying characteristic of the dark gauge boson.
As a concrete example, we revisit the misalignment \cite{Preskill:1982cy,Abbott:1982af,Dine:1982ah,Kim:1986ax} production of coherent oscillating dark gauge boson \cite{Nelson:2011sf}. 
Although the misalignment mechanism is widely used in the scalar case, especially in the axion dark matter production, the misalignment production of a massive vector is extremely difficult. It turned out that a naive model in Ref.~\cite{Nelson:2011sf} cannot produce the sizable relic vector boson because the energy density of the vector boson, unlike the scalar case, exponentially decreases during inflation\footnote{To have a correct relic dark matter density, the mass of a vector boson should be smaller than $10^{-38}$ GeV, but the Lyman-$\alpha$ forest gives a strong constraint on the relic dark matter density if the mass of a vector is less than $10^{-30}$ GeV. \cite{Graham:2015rva,Arias:2012az,Nakayama:2019rhg}}. Alternative models have been proposed to maintain the energy density of the vector boson during inflation. Refs. \cite{Arias:2012az,Alonso-Alvarez:2019ixv} suggested coupling of the vector boson to Ricci scalar, but it suffers from ghost instability of the longitudinal fluctuation \cite{Nakayama:2019rhg}. Refs. \cite{Nakayama:2019rhg,Nakayama:2020rka} suggested coupling to the inflaton, but isocurvature fluctuation and anisotropy of curvature fluctuation restrict the relic density of vector to be much smaller than the dark matter relic density \cite{Nakayama:2020rka}. 

In this paper, we suggest the misalignment mechanism for producing a sizable amount of the coherent vector boson may work if the vector boson mass increases by many orders of magnitude. We show this with the gauged quintessence model, where the quintessence scalar value and, thus, the vector boson mass increase greatly during the cosmic evolution. The dark gauge boson can achieve a comparable energy density as the CDM.

A brief review of the gauged quintessence model is in Sec.~\ref{Sec:gq}. In Sec.~\ref{Sec:dynamics}, we investigate the dynamics of the dark gauge boson from the inflation. In Sec.~\ref{Sec:constraint}, constraints on gauged quintessence model produced by the misalignment mechanism are discussed, before the summary and discussion in Sec.~\ref{Sec:Summary}.

\section{Gauged quintessence model}
\label{Sec:gq}

The gauged quintessence model \cite{Kaneta_2023} includes a complex scalar $\Phi$ and a $U(1)_\text{dark}$ gauge boson $\mathbb X_\mu$. The complex scalar is charged under the $U(1)_\text{dark}$ gauge symmetry, and consists of the radial part $\phi$ and the angular part $\eta$, $\Phi = \phi e^{i\eta}/\sqrt2$. The radial part $\phi$ is taken as the quintessence field. 

Under the unitary gauge,
\begin{equation}
    \eta= 0 , \quad X_{\mu} = \mathbb{X}_{\mu} +\frac{1}{g_X}\partial_{\mu}\eta \, ,
\end{equation}
the action of gauged quintessence model is given by\footnote{The FLRW metric with $g_{\mu\nu} = (-1,a^2,a^2,a^2)$ is used.}
\begin{equation}
\label{eq:actiongq}
    S \supset \int d^4x \; \sqrt{-g}\Big[-\frac{1}{2}(\partial_{\mu}\phi)(\partial^\mu \phi) -\frac{1}{4}X_{\mu\nu}X^{\mu\nu}-V_0(\phi) -\frac{1}{2}(g_X^{}\phi)^2 X_{\mu}X^{\mu} \Big],
\end{equation}
where $g_X$ is the dark gauge coupling constant. The last term in Eq.~\eqref{eq:actiongq} gives the mass of the dark gauge boson, and at the same time, it affects the quintessence dynamics. We call this term the gauge potential:
\begin{equation}
    V_{\text{gauge}}= \frac{1}{2}g_X^2\phi^2X_{\mu}X^{\mu}.
\end{equation}

For the scalar potential $V_0(\phi)$, we take the inverse-power potential (Ratra-Peebles potential) suggested by Ratra and Peebles \cite{PhysRevD.37.3406}:
\begin{equation}
V_0 (\phi) = \frac{M^{\alpha+4}}{\phi^{\alpha}} \, ,
\label{eq:RPpotential}
\end{equation}
where $\alpha>0$ and $M$ is chosen to fit the present-time dark energy density. This potential has a tracking behavior, so the wide range of initial conditions eventually converge to one common tracking solution \cite{Steinhardt:1999nw}. The tracking solution of the Ratra-Peebles potential gives the slow roll of the $\phi$ field, i.e. the kinetic energy of $\phi$ is much less than $V(\phi)$. This is essential for the quintessence to be the dark energy (See the reviews \cite{Martin:2008qp,Tsujikawa:2013fta}); likewise, the gauged quintessence model requires the slow roll of $\phi$ field.

The dynamics may alter when the quantum effects are considered \cite{Brax_2000,Doran:2002bc}. The 1-loop corrected potential of $\phi$ is given as\footnote{One may view the Ratra-Peebles potential as a quantum effective potential. However, it is unnatural to expect that couplings to the quintessence are tuned exactly to cancel the corrections from the gauge boson loops~\cite{Doran:2002bc}. } \cite{Kaneta_2023}
\begin{equation}
    V= V_0 + \frac{\alpha(\alpha+1) \Lambda^2}{32\pi^2} \frac{M^{\alpha+4}}{\phi^{\alpha+2}}+ \frac{(V_0^{\prime\prime})^2}{64\pi^2} \left(\ln\frac{V_0^{\prime\prime}}{\Lambda^2}-\frac{3}{2}\right) +\frac{3(m_{\text{X}}^2|_0)^2}{64\pi^2} \left(\ln\frac{m_{\text{X}}^2|_0}{\Lambda^2}-\frac{5}{6} \right),
    \label{eq:QV}
\end{equation} 
where $\Lambda$ is the cutoff scale, which we will take as the Planck mass ($M_{\text{Pl}}^{}\approx 1.22\times10^{19}~\text{GeV}$), and $^{\prime}$ denotes the partial derivative with respect to the $\phi$. Then the effective potential of $\phi$ is $V_{\text{eff}}=V+V_{\text{gauge}}$. Following the discussions in Ref.~\cite{Kaneta_2023}, we find the third term is negligibly small, and the last term gives an upper bound on $m_X^{}$ as $m_X^{}\lesssim 10^{-11}$ GeV.

One important feature of the gauged quintessence model is the evolution of $m_X^{}$. (For some other examples of the mass-varying particles, see Refs.~\cite{Casas:1991ky,Garcia-Bellido:1992xlz,Anderson:1997un,Fardon:2003eh,Berlin:2016bdv,Davoudiasl:2019xeb,Boubekeur:2023fqo,ChoeJo:2023ffp}.) The masses of $\phi$ and $X$ are given by 
\begin{equation}
m_{\phi}^2 = \frac{\partial^2 V_{\text{eff}}}{\partial \phi^2} , \quad m_X^2 = g_X^2 \phi^2,
\label{eq:treemass}
\end{equation} 
and $m_X^{}$ evolves when $m^{}_{\phi}\gtrsim H$. As we will demonstrate in Sec.~\ref{Sec:dynamics}, the $X$ behaves as a non-relativistic matter when $m_X^{} \gtrsim H$. However, the dark gauge boson cannot be the candidate of sole dark matter since the mass of the $X$ can change many orders of magnitude after the matter-radiation equality. So we assume an additional species that gives the dominant contribution to the dark matter relic density. The $X$ density can be amplified by the mass-increasing effect, and this property will be crucial in the production of $X$ as we will demonstrate in Sec.~\ref{Sec:Subdominant during the inflation}.

\section{Dynamics of the gauged quintessence model}
\label{Sec:dynamics}

There are two distinct contributions to the coherent dark gauge boson after inflation. One is the production from the misalignment, which originates from the homogeneous mode at the beginning of the inflation. The wavelength of this mode is much larger than the current observable scale of the universe, so it remains homogeneous until today. The other contribution is the production from the quantum fluctuation (gravitational production). These fluctuations can be exponentially stretched during inflation, and some of the long wavelength fluctuations could remain homogeneous until today. \footnote{A more detailed discrimination of the two, see, e.g., Ref. \cite{Kaneta:2023kfv}.}
We will consider the case that only the misalignment production of the $X$ gauge boson is effective; namely, the other contribution is subdominant.

In this section, we discuss the dynamics of a homogeneous mode of $X$ and $\phi$ field. From this, one can calculate various quantities such as mass and energy density. In brief, the dynamics of the $X$ field and $\phi$ field are determined by the hierarchy between the mass of each field and the Hubble parameter. In other words, the dynamics of $X$ or $\phi$ field changes when $H \sim m^{}_X$ or $H \sim m^{}_\phi$, respectively. To simplify the discussion, we define $a_\text{tr}$ $(a_\text{nr})$ as the scale factor when $H\sim m^{}_{\phi}$ ($H\sim m^{}_X$). The $\phi$ field follows the tracking solution after the $a_{\text{tr}}$; and $X$ boson starts coherent oscillation after the $a_{\text{nr}}$, and it behaves as a non-relativistic matter.
Also, we denote $a_\text{ini}$, $a_\text{end}$, and $a_\text{eq}$ as the scale factor when inflationary, radiation-dominated, and matter-dominated epoch begins.\footnote{We assumed instant reheating to simplify discussions. Therefore, $a_\text{end}$ corresponds to both the end of inflation and the beginning of the radiation-dominated epoch.}

\subsection{Dynamics of the dark gauge boson}

The equation of motion of dark gauge boson is obtained from the action (Eq.~\eqref{eq:actiongq}) as
\begin{equation}
\begin{split}    
    \partial_\mu \left(\sqrt{-g}  X^{\mu\nu}\right) - \sqrt{-g}m_X^{2} X^\nu = 0 .
\end{split}
\end{equation}
For the homogeneous mode $X_{\mu}(t,\vec x) = X_{\mu}(t)=(X_0(t), \vec{X}(t))$, and it gives
\begin{equation}
X_0 =0 ,\quad
 \ddot{\vec X}+H\dot{\vec X} +m_X^{2}\vec X=0  .
\label{eq:hom_eom}
\end{equation}
We set our coordinate such that $\vec{X} =(0,0,X)$ without the loss of generality, so the energy density of the homogeneous mode is written as
\begin{equation}
\rho_X^{} = \rho_X^{(k)} + \rho_X^{(v)} \quad \text{where}  \quad \rho_X^{(k)} = \frac{1}{2a^2} \dot X^2 \quad \text{and} \quad \rho_X^{(v)} = \frac{1}{2a^2} 
 m_X^{2} X^2 .
\label{eq:Xdensity}
\end{equation}

In order to find the evolution of the $\rho_X$, one needs to solve the equation of motion (Eq.~\eqref{eq:hom_eom}). To do so, we use the physical field defined as $\overline{X}=X/a$, and rewrite Eq.~\eqref{eq:hom_eom} as
\begin{equation}
    \ddot{\overline{X}} +3H\dot{\overline{X}} +\left(m_X^{2}+\frac{1-3w_b}{2}H^2 \right)\overline{X} = 0 ,
    \label{eq:red_hom_eom}
\end{equation}
where $w_b$\footnote{$w_b=-1$ in the inflationary era, $w_b=1/3$ in the radiation dominated era, and $w_b=0$ in the matter dominated era.} is the background equation of state. This equation is analogous to the equation of motion for the scalar field, except for the Hubble-induced mass term.

\begin{figure}[t]
\begin{subfigure}{0.5\textwidth}
\centering
\includegraphics[width=0.99 \linewidth]{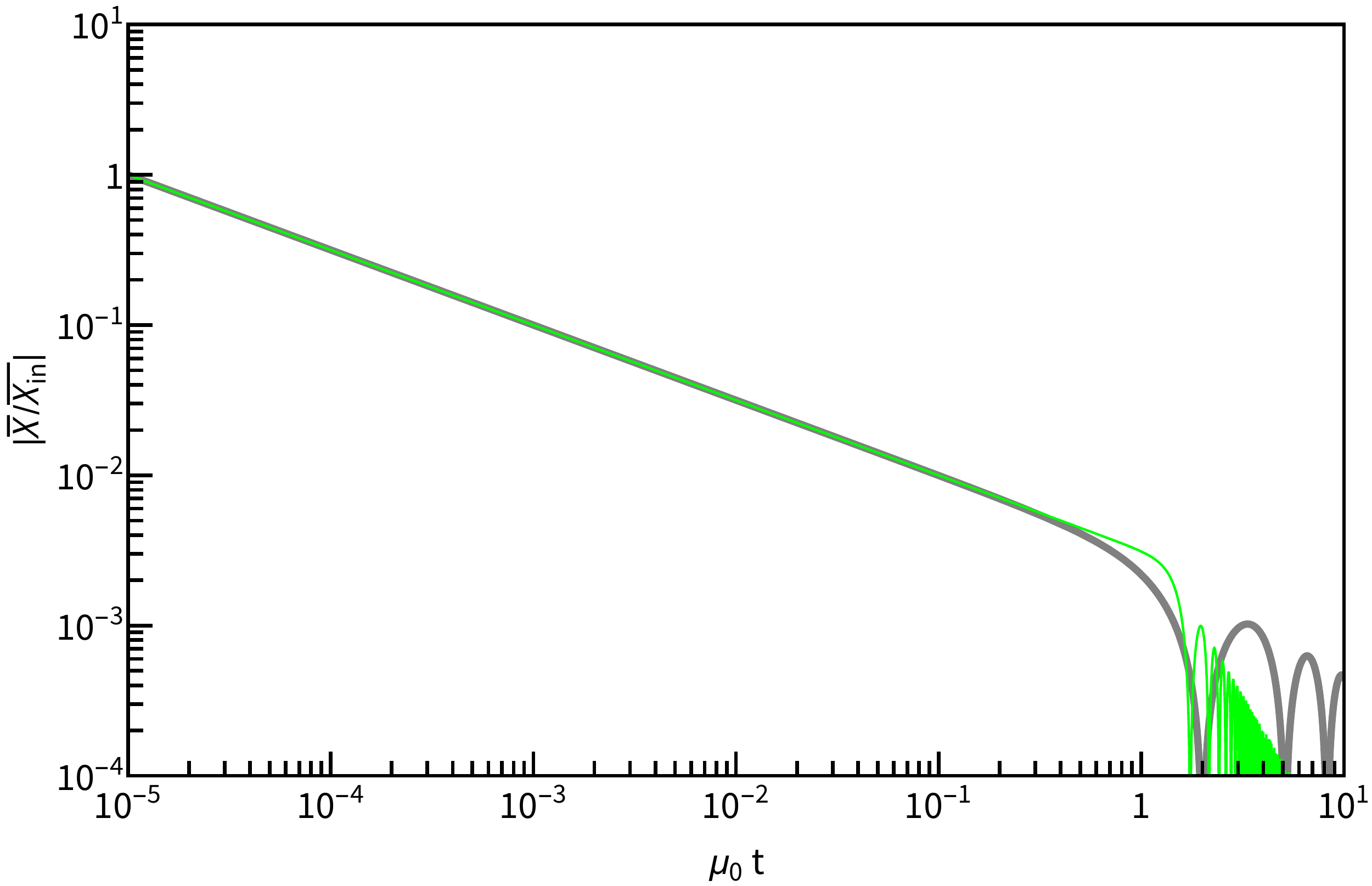}
\caption{}
\label{fig:Xevolve_sub-1}
\end{subfigure}
\begin{subfigure}{0.5\textwidth}
\centering
\includegraphics[width=0.99\linewidth]{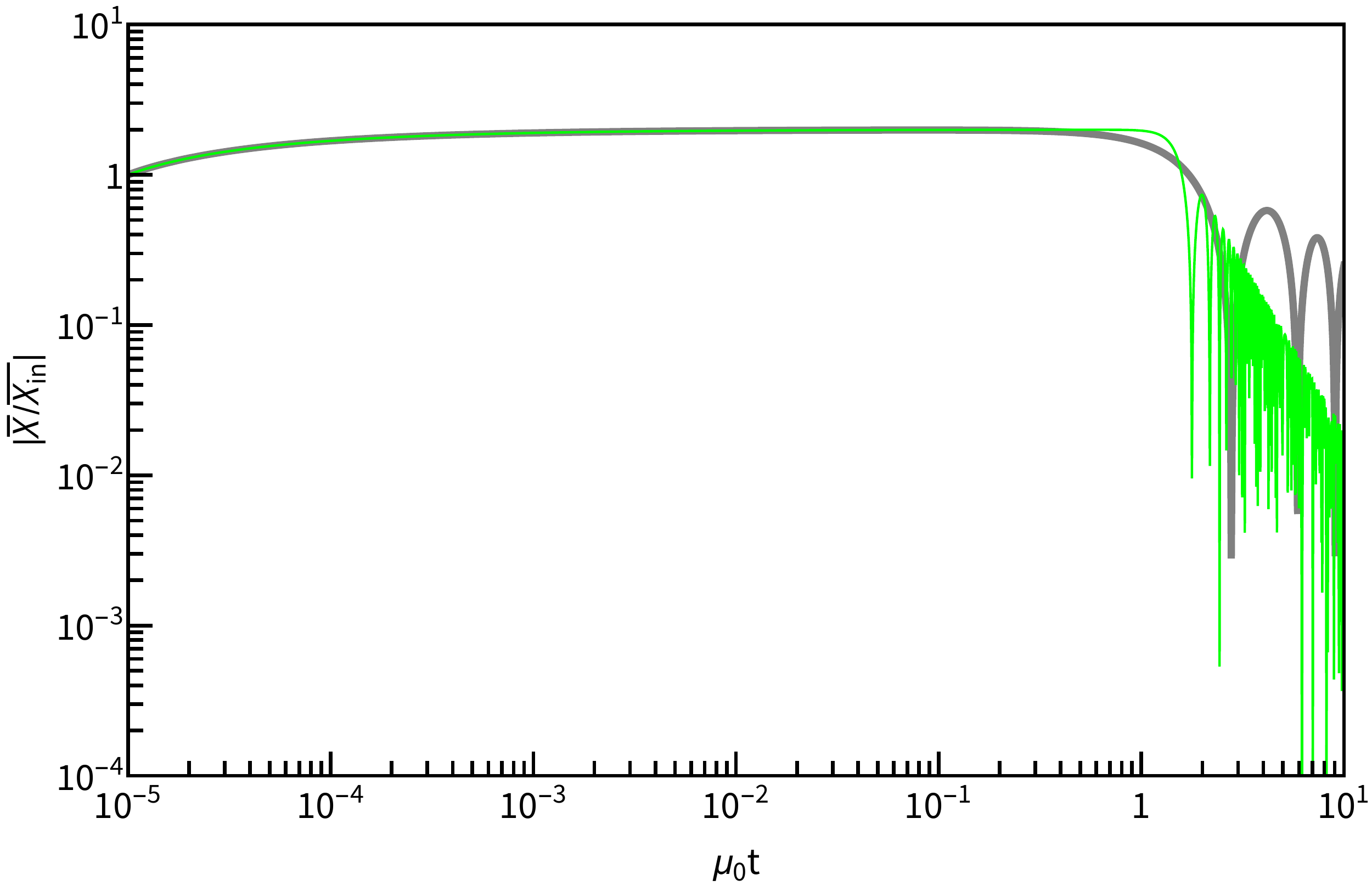}
\caption{}
\label{fig:Xevolve_sub-2}
\end{subfigure}
    \caption{Evolution of the dark gauge boson field normalized by the initial field value $\overline{X_{\text{in}}}$ during the radiation-dominated era.  The gray curve has a constant mass $m_X^{} = \mu_0$, and the green curve has time varying mass $m_X^{}=\mu_0( \mu_0 t)^3$. The numerical evaluation starts at $\mu_0 t= 10^{-5}$. (a) For the potential energy-dominated case, $\overline{X} \propto 1/\sqrt{t} \propto 1/a$. (b) For the kinetic energy-dominated case, $\overline{X}$ is almost constant, except for the $\mathcal{O}(1)$ modification.
    For both cases, the evolution of the $\overline{X}$ until $H\sim m_X^{}$ is not affected by the $m^{}_X$. The bottoms of oscillations reach 0, although they do not appear due to the resolution of the plots.}  
    \label{fig:Xevolve}
\end{figure}

The solution of this equation when $m_X^{} \ll H$ is \cite{Nakayama:2019rhg}
\begin{equation}
    \overline{X} = c_1 a^{-1} +c_2 a^{(3w_b-1)/2} \quad\Leftrightarrow\quad X = c_1  +c_2 a^{(3w_b+1)/2} ,
    \label{eq:Xsol_aftinf}
\end{equation}
where the $c_1$ and $c_2$ are undetermined coefficients, which can be determined by the initial condition. The $c_1$ solution corresponds to the potential energy dominated initial condition ($\rho_X^{(k)}\ll \rho_X^{(v)}$), and $c_2$ solution corresponds to the kinetic energy dominated initial condition ($\rho_X^{(k)}\gg \rho_X^{(v)}$).
These solutions are valid even when $w_b=1/3$. We give some numerical examples of $\overline{X}$ evolution during the radiation-dominated era in Fig.~\ref{fig:Xevolve}. 
Then the evolution of the $X$ density when $m_X \ll H$ is given as
\begin{equation}
    \begin{dcases}
        \rho_X =\frac{1}{2}\left(\left(\frac{3w_b+1}{2}\right)^2H^2+m_X^2 \right) \overline X^2\propto a^{-4}   & \text{for } \rho_X^{(k)}\gg\rho_X^{(v)},\\ \rho_X = \frac{1}{2}m_X^{2} \overline X^2 \propto m_X^2a^{-2}   & \text{for } \rho_X^{(k)}\ll\rho_X^{(v)}.
    \end{dcases}
    \label{eq:aftinf_initial}
\end{equation}
We wrote $m_X^2 a^{-2}$ instead of $a^{-2}$ to explicitly show the mass varying effect. 

On the other hand, the evolution of $X$ when $m_X^{} \gtrsim H$ is found in the adiabatic limit as \cite{Wentzel1926,Kramers1926,Brillouin:1926blg}
\begin{equation}
    \overline{X}\propto \text{Re} \left[ \frac{1}{\sqrt{a^3m_X^{}}}e^{i\int dt\, m_X^{}}\right] \quad \Leftrightarrow\quad X \propto \text{Re} \left[ \frac{1}{\sqrt{a m_X^{}}} e^{i\int dt\, m_X^{}}\right],
\end{equation}
where the $\text{Re}[f]$ denotes the real part of $f$.
Hence, the kinetic energy and potential energy of the $X$ averaged over the oscillations are the same in this regime. Thus the energy density evolves as
\begin{equation}
    \rho_X \propto m_X^{} a^{-3} ,
\end{equation}
which is the same as the evolution of the non-relativistic particle. We summarize the results in Tab.~\ref{tab:evolution}.

\renewcommand{\arraystretch}{1.5}
\begin{table}[t]
    \centering
    \begin{tabular}{| c | c | c |}
  \hline
    Initial condition   &  $H \gg m_X^{}$  & $H \lesssim m_X^{}$\\
    \hline 
        $\rho_X^{(k)} \gg \rho_X^{(v)}$  & $\rho_X^{} \propto a^{-4}$ & $\rho_X^{} \propto m_X^{} a^{-3}$\\
        \hline
     $\rho_X^{(k)} \ll \rho_X^{(v)}$ &  $\rho_X^{} \propto m_X^2  a^{-2}$ & $\rho_X^{} \propto m_X^{} a^{-3}$\\\hline
    \end{tabular}
    \caption{Evolution of $\rho_X$ for different initial conditions. For both cases, the energy density evolves as a non-relativistic species after $H \sim m_X^{}$.}
    \label{tab:evolution}
\end{table}
Interestingly, the energy density of the dark gauge boson can be largely enhanced if the mass of $X$ increases during cosmic evolution. Also, this effect is much more noticeable for the potential energy-dominated initial condition. Thus, the dark gauge boson energy density could even be comparable to the CDM energy density despite the exponential suppression during inflation. 

Although there is a mass-varying effect, it is hard to achieve a sufficient relic density of $X$ in the kinetic energy-dominated scenario due to $a^{-4}$ suppression. Assuming that $X$ becomes non-relativistic after inflation, the present energy density of $X$ is at most $10^{-50}$ times that of matter. There is a possibility that $X$ becomes non-relativistic during inflation but the kinetic energy-dominated case cannot provide sufficient energy in this case, too. Therefore, we will consider the potential energy-dominated case only, hereafter.

\subsection{Dynamics of the quintessence}
\label{sec:Dq}
The equation of motion for the homogeneous mode of $\phi$ is given as 
\begin{equation}
    \ddot \phi + 3 H \dot \phi + \frac{\partial V_{\text{eff}}}{\partial \phi} = 0.
    \label{eq:homEomPhi}
\end{equation}
The dominant terms of $V_{\rm eff}$ are\footnote{The tree-level potential $V_0$ dominates only when $\phi > \sqrt{\alpha(\alpha+1)\Lambda^2/(32\pi^2)} \sim 0.1 \Lambda$. If we take $\Lambda=M_{\text{Pl}}$, the first term in Eq.~(\ref{eq:V1}) mostly governs the dynamics over the tree-level potential except for a short period near present \cite{Kaneta_2023}.  }
\begin{equation}
    V_\text{eff} \approx \frac{\alpha(\alpha+1) \Lambda^2}{32\pi^2} \frac{M^{\alpha+4}}{\phi^{\alpha+2}} + \frac{1}{2a^2}g_X^2 X^2 \phi^2.
    \label{eq:V1}
\end{equation}
So, the dynamical minimum of the potential is generated from the combination of the inverse power term and a quadratic term.  The $\phi$ field and mass of $\phi$ at the minimum of $V_{\text{eff}}$ are given as 
\begin{equation}
\begin{split}
     \phi_\text{min} &= \left(\frac{\alpha(\alpha+1)(\alpha+2)\Lambda^2 M^{\alpha+4}}{32\pi^2 g_X ^2\overline X^2}\right)^{1/(\alpha+4)} \propto \overline X ^{-2/(\alpha+4)},\\
        m_\phi(\phi_{\text{min}}) &= \sqrt{\alpha+4}  g_X \overline X   \propto \overline X .
    \label{eq:phimin} 
 \end{split}
\end{equation}

The dynamics of quintessence is determined by the relative size of $m_\phi$ and $H$, and the magnitude of $ V_{\text{gauge}}^{\prime\prime}$. If $H\gg m_\phi$, $\phi$ becomes constant due to the Hubble friction, and so the $m_{\phi}$ is fixed, too. On the other hand, if $H \lesssim m_\phi$, then the $\phi$ field rolls down the potential. Two different situations can happen in this case. If the $V_{\text{gauge}}^{\prime\prime} \gtrsim H^2$, the potential is sufficiently steep about the minimum so the $\phi$ field follows the minimum of the potential. Thus, we approximate that $\phi \sim \phi_{\text{min}}$. However, if $V_{\text{gauge}}^{\prime\prime}\ll H^2$, $\phi$ field cannot reach the potential minimum during the Hubble time, and $V_{\text{gauge}}$ can be ignored. In this case, the $\phi$ field rolls down until $m_{\phi} \sim H$, and follows the tracking solution of the inverse power potential. During the tracking, the scaling of $\phi$ and $m_\phi$ becomes \cite{Zlatev:1998tr}
\begin{equation}
\phi \propto a^{\frac{3(w_b+1)}{\alpha+4}}, \qquad m_\phi \propto a^{-\frac{3(w_b+1)}{2}}.
\end{equation}
Hence, the scaling behavior of $m_\phi$ is the same as the $H$, and the $\phi\sim \text{constant}$ during inflation. 

\begin{figure}[tb]
\centering
\includegraphics[width=0.7\linewidth]{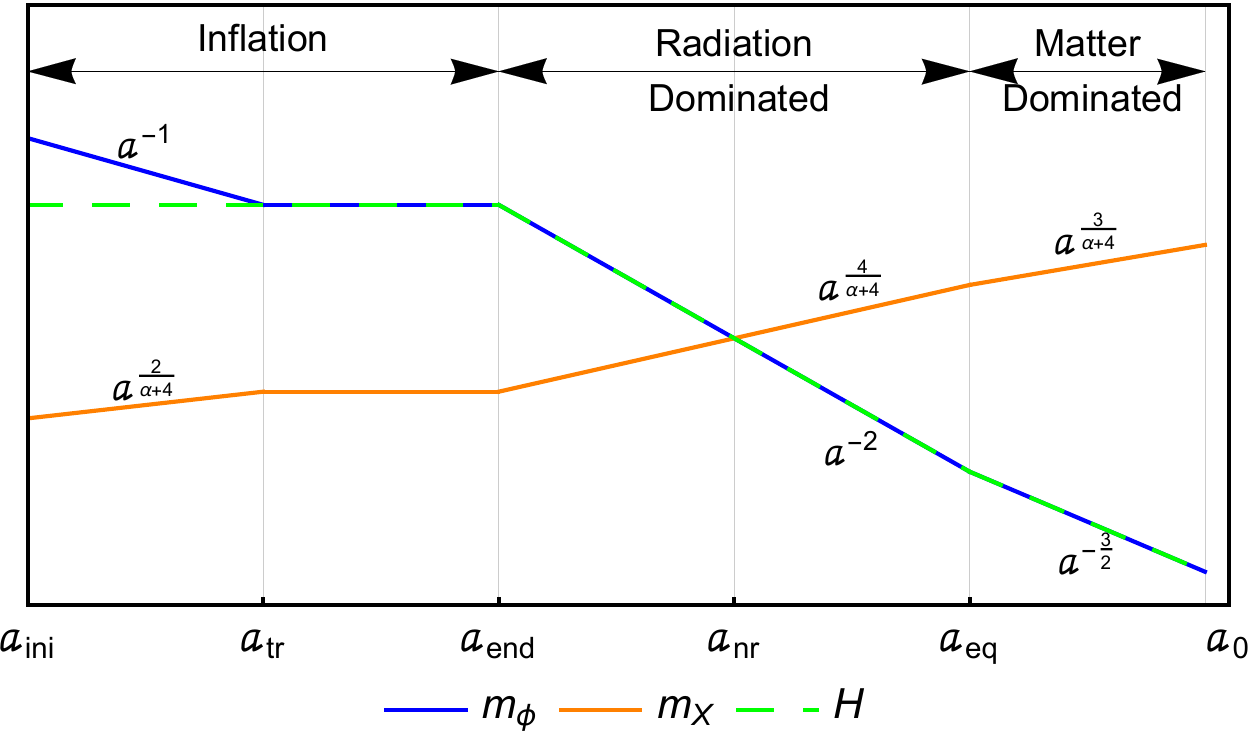}
\label{fig:scenario-2}

    \caption{Schematic description of scaling behavior of $m_\phi$, $m_X$ and $H$ for the scenario in Sec. \ref{sec:Dq}. Tracking behavior begins at $a_\text{tr}$ when $m_\phi \sim H$ is achieved. In this scenario, it is assumed that $a_\text{tr}$ is in the inflationary era. $a_\text{nr}$, when $m_X \sim H$ is in the radiation dominated era. This figure is a rough description and the relative scales are not exact. The blank region before the $a_0$ corresponds to the dark energy dominated era with the tree-level potential domination. This era is extremely short compared to other eras.}  
    \label{fig:scenario}
\end{figure}

Let us discuss a viable scenario to produce sufficient relic $X$ density. We consider that $m_{\phi}$ is much larger than $H_{\text{inf}}$ at the beginning of inflation but eventually becomes comparable to $H_{\text{inf}}$ before inflation was terminated\footnote{Although the scenarios such that $m_\phi \ll H_\text{inf}$ or $m_\phi \gg H_\text{inf}$ during the entire inflation era are possible, they cannot provide sufficient relic density of $X$. Therefore, we will not deal with these scenarios in the main text. Refer to Appendix~\ref{Sec:Other scenarios}.}. Therefore, the $\phi$ field initially moves along the potential minimum and joins the tracking solution before the  $a_{\text{end}}$. 
Schematic description of $m_\phi$, $m_X$, and $H$ of this scenario is given in Fig. \ref{fig:scenario}. Using the dynamics of $\overline X$ and $\phi$ discussed in the earlier part of this section, one can obtain the scaling behavior of $\rho_X$ and $\rho_{\phi}$. The dynamics of $\overline{X}$, $\phi$ ($m^{}_{X}$), $\rho^{}_{X}$ and $\rho_{\phi}$ are summarized in the following texts and Tab.~\ref{tab:sc2}. For simplicity, we take that the $V_{\text{gauge}}$ is negligible after $a_{\text{tr}}$. Also, we omit the short dark energy or tree-level potential dominated epoch right before the $a_0$. This simplification does not alter the order of magnitude estimation of the relic densities of $X$, and $\phi$.

	\begin{table}[h!]
		\centering
		\aboverulesep=0ex
		\belowrulesep=0ex
		\begin{tabular}{|c|c|c|c|c|}
			\midrule\rule{0EM}{1.1EM}
			& $\overline X$ & $\phi$ & {$\rho^{}_X$} & $\rho_\phi$ \\
			\midrule\rule{0EM}{1.4EM}
			$ a_\text{ini}<a<a_\text{tr} $ & $\propto a^{-1}$& $\propto a^{\frac{2}{\alpha+4}}$  &  $\propto a^{-\frac{2\alpha+4}{\alpha+4}}$ & $\propto a^{- \frac{2\alpha+4}{\alpha+4}}$  \\
			\midrule  \rule{0EM}{1.4EM}  $a_\text{tr}<a<a_\text{end}$ & $\propto a^{-1}$  & $\sim$ constant        & $\propto a^{-2}$ & $\sim$ constant  \\
			\midrule  \rule{0EM}{1.4EM}  $a_\text{end}<a<a_\text{nr}$ & $\propto a^{-1}$ & $\propto a^{\frac{4}{\alpha+4}}$ &        $\propto a^{-\frac{2\alpha}{\alpha+4}}$ & $\propto a^{- \frac{4\alpha+8}{\alpha+4}}$ \\
			\midrule\rule{0EM}{1.4EM}
			$a_\text{nr}<a<a_\text{eq}$ & $\propto a^{-\frac{3\alpha+16}{2\alpha+8}}$ & $\propto a^{\frac{4}{\alpha+4}}$ &  $\propto a^{-\frac{3\alpha+8}{\alpha+4}}$ & $\propto a^{- \frac{4\alpha+8}{\alpha+4}}$ \\
			\midrule  $a_\text{eq}<a<a_0$ & $\propto a^{-\frac{3\alpha+15}{2\alpha+8}}$ & $\propto a^{\frac{3}{\alpha+4}}$  & $\propto a^{-\frac{3\alpha+9}{\alpha+4}}$ & $\propto a^{- \frac{3\alpha+6}{\alpha+4}}$ \\
			\midrule
		\end{tabular}%
      \caption{Behaviors of $\overline X$, $\phi$, $\rho^{}_X$ and $\rho_\phi$ when $m_\phi > H_\text{inf}$ at the beginning of inflation and $m_\phi \sim H_\text{inf}$ is achieved during inflation.}
      \label{tab:sc2}
	\end{table}%

Since we have determined the scaling behaviors, we can calculate various quantities if we know the $a_{\text{ini}}$, $a_{\text{tr}}$, $a_{\text{end}}$, $a_{\text{nr}}$. For simplicity, we assume the instant reheating\footnote{Under the assumption of the instant reheating, we may estimate the reheating temperature $T_\text{rh}$ via $\rho_\text{inf}=\rho_\text{rad}$ where $\rho_\text{rad}=(g_*(T_\text{rh})\pi^2/30)T_\text{rh}^4/m_\text{Pl}^2$ with $g_*(T)$ the effective number of relativistic degrees of freedom at $T$, and  $m^{}_\text{Pl} = M^{}_\text{Pl}/\sqrt{8\pi}$.}, so the $a_{\text{end}}$ is given from the following relation as
\begin{equation}
    H_\text{inf} \sim \sqrt{\Omega_r}H_0 \left(\frac{a_0}{a_\text{end}}\right)^{2}.
    \label{eq:aend}
\end{equation}
Also, we take the number of e-folds of inflation as $N=60$. This gives $a_{\text{ini}}=e^{-60}a_{\text{end}}$. Let us determine the $a_{\text{tr}}$ and $a_{\text{nr}}$. The Eqs.~\eqref{eq:aftinf_initial} and \eqref{eq:phimin} gives
 \begin{equation}
    m_\phi \sim \frac{\sqrt{2(\alpha+4)\rho_X}}{\phi} \sim \sqrt{2(\alpha+4)} \rho_X^{\frac{\alpha+4}{2\alpha+4}} \left(\frac{\alpha(\alpha+1)(\alpha+2)\Lambda^2 M^{\alpha+4}}{64\pi^2 }\right)^{-\frac1{\alpha+2}} \quad \text{for }a\lesssim a_{\text{tr}}.
 \end{equation}
Using that $m_\phi \propto a^{-1}$ for $a<a_{\text{tr}}$, and $m_{\phi}\sim H$ at $a_{\text{tr}}$,
 \begin{equation}
     \frac{a_{\text{tr}}}{a_\text{ini}} \sim  \sqrt{2(\alpha+4)} \rho_X(a_\text{ini})^{\frac{\alpha+4}{2\alpha+4}}  \left(\frac{\alpha(\alpha+1)(\alpha+2)\Lambda^2 M^{\alpha+4}}{64\pi^2 }\right)^{-\frac1{\alpha+2}} H_\text{inf}^{-1} .\label{eq:atr}
 \end{equation}
 Also, $\phi$ at $a_{\text{tr}}$ is given as
\begin{equation}
\phi(a_{\text{tr}}) \sim  \left(\frac{\alpha(\alpha+1)(\alpha+2)(\alpha+4) \Lambda^2 M^{\alpha+4}}{32\pi^2 H_\text{inf}^2}\right)^{\frac1{\alpha+4}} .
\end{equation}
The value of $\phi$ remains constant during the remaining inflationary era. During the radiation domination era, $\phi \propto a^{\frac{4}{\alpha+4}}$, and $H \propto a^{-2}$. Since $H\sim m^{}_X$ at $a_\text{nr}$, we have
\begin{equation}
m^{}_X(a_\text{nr}) \sim g^{}_X \phi(a_\text{tr}) \left(\frac{a_\text{nr}}{a_\text{end}}\right)^{\frac4{\alpha+4}} \sim H_\text{inf}  \left(\frac{a_\text{nr}}{a_\text{end}}\right)^{-2}.\end{equation}
This relation gives $a_\text{nr}/a_\text{end}$ as
\begin{equation}
\frac{a_\text{nr}}{a_\text{end}} \sim g_X^{- \frac{\alpha+4}{2\alpha+12}}H_\text{inf}^\frac12\left(\frac{\alpha(\alpha+1)(\alpha+2)(\alpha+4) \Lambda^2 M^{\alpha+4}}{32\pi^2 }\right)^{- \frac{1}{2\alpha+12}} .
\label{eq:anr}
\end{equation}
Therefore, we can determine all relevant scale factors required to calculate various quantities for this scenario.

\section{Constraints}\label{Sec:constraint}

In this section, we show various constraints on the scenario in the previous section, especially for $\alpha=1$. We set $M=2.2\times 10^{-6}$ GeV to reproduce the present dark energy density. 

\subsection{Backreaction of $\rho_X^{}$ during the inflation}\label{Sec:Subdominant during the inflation}

In order to have a successful inflation, the energy density of $X$ should be much smaller than the inflaton energy density. Since the energy density of $X$ monotonically decreases during inflation, the energy density of $X$ at the start of inflation has an upper limit by 
\begin{equation}
    \rho_{X}(a_\text{ini}) \ll \rho_{\text{inf}}= 3m_{\text{Pl}}^2H_{\text{inf}}^2,
    \label{eq:const_inf}
\end{equation}
where $m_\text{Pl} = M_\text{Pl}/\sqrt{8\pi}\simeq 2.4\times10^{18}$ GeV is the reduced Planck mass. Here, the $\rho_{\text{inf}}$ is the energy density of inflaton. If we assume that the $\phi$ settles down close to the potential minimum, we have $\rho_{\phi}\sim \rho_{X}$. So the backreaction from the $\phi$ can be also verified by Eq.~\eqref{eq:const_inf}.

The quantitative constraint on the main scenario is the following. Using Tab.~\ref{tab:sc2}, we have
\begin{equation}
    	\rho^{}_X(a_\text{ini})=\rho^{}_X(a_0)  \left(\frac{a_\text{eq}}{a_0}\right)^{-12/5} \bigg(\frac{a_\text{nr}}{a_\text{eq}}\bigg)^{-11/5} \bigg(\frac{a_\text{end}}{a_\text{nr}}\bigg)^{-2/5}\bigg(\frac{a_\text{tr}}{a_\text{end}}\bigg)^{-2}  \bigg(\frac{a_\text{ini}}{a_\text{tr}}\bigg)^{-6/5},
\end{equation}
where $a_\text{tr}$ and $a_\text{nr}$ can be calculated from Eqs.~\eqref{eq:atr} and \eqref{eq:anr}.  
When we assume $\ln(a_\text{end}/a_\text{ini})=60$, for example, the constraint on $\rho_{X}(a_{\text{ini}})$ to avoid the backreaction is written as 
\begin{equation}
\frac{\rho^{}_{X}(a_{\text{ini}})}{3m_{\text{Pl}}^2 H_{\text{inf}}^2}= 3.5 \times 10^{-17} \left(\frac{\rho_{X}(a_0)}{\rho_{\text{CDM}}(a_0)}\right)^{3/5}\left(\frac{H_{\text{inf}}}{10_{}\, \text{GeV}}\right)^{-7/5}\left(\frac{g^{}_X}{10_{}^{-31}}\right)^{27/70} < 1.
\end{equation}
For a fixed amount of $\rho_X^{}$, we have a lower bound on $H_{\text{inf}}$ or an upper bound on $g_X^{}$.

\subsection{Isocurvature fluctuation}

The fluctuation of the $X$ field is an independent degree of freedom from the inflaton fluctuation. Therefore, the isocurvature fluctuation is generated from the quantum fluctuation of the $X$. Since the production of the transverse modes is suppressed for the minimal vector field \cite{Graham:2015rva}, only the longitudinal modes contribute to the fluctuation of the $X$. The power spectrum of the longitudinal mode is defined as
\begin{equation}
 \quad \langle \overline{X}_L(\vec{k}_1) \overline{X}_L^{\ast}(\vec{k}_2) \rangle \equiv \frac{2\pi^2}{k^3} \mathcal{P}_{\overline{X}_L}(\vec{k}_1) (2\pi)^3\delta^{(3)}(\vec{k}_1-\vec{k}_2),
\label{eq:powerSpDef}
\end{equation} 
and one can calculate that (see App.~\ref{Sec:Growth of the fluctuations} for the detail)
\begin{equation}
    \mathcal{P}_{\overline{X}_L}(\vec{k},a_{\text{end}}) = \zeta \left(\frac{k H_{\text{inf}}}{2\pi a_{\text{end}}m_X^{}(a_{\text{end}})}\right)^2,
    \quad\zeta =\begin{dcases}
        1 &\; \text{for} \; a_{\text{tr}}< \frac{k}{H_{\text{inf}}},\\ \frac{2^{14/5}}{\pi} \Gamma^2\left(\frac{19}{10}\right) \left(\frac{a_{\text{tr}}H_{\text{inf}}}{k}\right)^{4/5} &\; \text{for} \; a_{\text{tr}}> \frac{k}{H_{\text{inf}}}.
    \end{dcases}
    \label{eq:LmodeMain}
\end{equation}
Here, the power spectrum is proportional to the inverse of $m_{X}$, so the isocurvature constraint is stronger for the smaller $m_X^{}$. 

 Then, the isocurvature fluctuation is given by\footnote{We assume that the dark gauge boson constitutes a fraction of total cold dark matter, and denote the other CDM component as $\rho_{\text{CDM}}$. If the $\rho_X^{}$ is comparable to the $\rho_{\text{CDM}}$, the CMB fitting parameters may change. But, we will assume that this does not significantly affect the order of magnitude constraints given in this section.} \cite{Wands:2000dp,Kawasaki:2008sn,Langlois:2008vk}
\begin{equation}
\begin{split}
       S_{\text{CDM}}^{}(\vec{k},a_{\text{lss}}) &\sim \frac{\delta \rho_X^{}(\vec{k},a_{\text{lss}})}{\rho_X^{}(a_{\text{lss}})+\rho^{}_{\text{CDM}}(a_{\text{lss}})}  \sim \frac{2m^{2}_X(a_{\text{lss}}) \overline{X}(a_{\text{lss}})\delta\overline{X}(k,a_{\text{lss}})} {m^2_X(a_{\text{lss}}) \overline{X}^2(a_{\text{lss}})+2\rho^{}_{\text{CDM}}(a_{\text{lss}}) }\\ &\sim \frac{2\sqrt{\mathcal{P}_{\overline{X}_L}(\vec{k},a_{\text{end}})}/\overline{X}(a_{\text{end}})}{1+\rho^{}_{\text{CDM}}(a_{\text{lss}})/\rho^{}_X(a_{\text{lss}}) },
\end{split}
\end{equation}
where the $\delta\rho_{X}$ is evaluated at the uniform density slicing \cite{Lyth:2004gb}, and the $a_{\text{lss}}$ is the scale factor at last scattering surface. In the last step, we used that the evolution of the super horizon mode and the zero modes are the same after the inflation. (See, App.~\ref{Sec:Growth of the fluctuations}.) 

The $\overline{X}(a_{\text{end}})$ can be calculated as\footnote{This implies that the $X/a$ may have a super Planckian value for $a \ll a_{\text{nr}}$. We will accept this possibility and suppose that UV physics above the Planck scale does not significantly affect the following discussion.}
\begin{equation}
    \overline{X}(a_{\text{end}}) \sim 5.4\times 10^{21}\,\text{GeV}\, \left(\frac{\rho_X(a_0)}{\rho_{\text{CDM}}}\right)^{1/2} \left(\frac{H_{\text{inf}}}{10\,\text{GeV}}\right)^{1/2} \left(\frac{g^{}_X}{10_{}^{-31}}\right)^{-19/28}.
\end{equation}
Also, from the conservation of the number density of the dark gauge boson, i.e. $\rho_X^{0}/m_X^{0}=a^3\rho_X^{}/m_X^{}$, we have
\begin{equation}
    \frac{\rho^{}_{\text{CDM}}(a_{\text{lss}})}{\rho^{}_X(a_{\text{lss}})} \sim 67\, \frac{\rho^{}_{\text{CDM}}(a_{0})}{\rho^{}_X(a_{0})}.
\end{equation}
Then, the isocurvature constraint can be calculated from the $S_{\text{CDM}} \lesssim 9\times 10^{-6}$ at $k\sim0.05$ Mpc$^{-1}$ \cite{Planck:2018jri}\footnote{The CMB power spectrum probes $k= 0.002 - 0.1$ Mpc$^{-1}$. In the mass unit, this corresponds to $k=10^{-41} - 7\times 10^{-40}$ GeV.}. We show this constraint in figure.~\ref{fig:constraint}.

\subsection{Stochastic correction to the quintessence evolution}

So far, we assumed that the homogeneous mode of the $\phi$ field follows the trajectory determined by the equation of motion (Eq.~\eqref{eq:homEomPhi}). Actually, this is not always true since the quantum fluctuations of the quintessence field can affect the motion of the quintessence field \cite{Malquarti:2002bh,Martin:2004ba}. During the inflation, fluctuations of $\phi$ field can exit the horizon (i.e. $k/a < H_{\text{inf}}$), and gives a random jump on the homogeneous part of the $\phi$ field \cite{Malquarti:2002bh}. Therefore, the evolution of the homogeneous part of the $\phi$ field is under a stochastic process \cite{starobinsky1984fundamental,starobinsky1986field}. However, we will demonstrate that the stochastic effect is negligible for the scenario in Sec.~\ref{Sec:dynamics}.

Let us split the quintessence field into the homogeneous part\footnote{Formally, the homogeneous part is a collection of long wavelength mode with $k<aH$ \cite{starobinsky1986field}.} ($\phi_0$) and inhomogeneous perturbation ($\delta \phi$) such that $\phi=\phi_0+\delta \phi$. 
The evolution of the Fourier component with momentum $k$ is described by the Klein-Gordon equation \cite{Malquarti:2002bh} as
\begin{equation}
    \ddot{\tilde{\phi}}_k(t)+3H\dot{\tilde{\phi}}_k(t) +\left((k/a)^2+ m_{\phi}^2\right) \tilde{\phi}_k(t)=0,
    \label{eq:kmodeEOM}
\end{equation}
where the $\tilde{\phi}_k(t)$ is the fluctuation with a wave number k. For the nearly massless case, the power spectrum of the fluctuation is given by $P_{\phi}(k,t) \approx (H/(2\pi))^2$. This implies that the size of the random jump of the $\phi$ field per Hubble time is about $H/(2\pi)$ \cite{Malquarti:2002bh}. On the other hand, compared to the nearly massless case, a magnitude of fluctuation is suppressed when $m_{\phi}\gtrsim H$. 

In the scenario in Sec. \ref{Sec:dynamics}, the $\phi$ field is initially trapped in the potential minimum with $m_{\phi}$ larger than $H$. Thus, we can approximate that the $\phi$ follows the potential minimum, and the stochastic effect is negligible. After the $a_{\text{tr}}$, the $V_{\text{gauge}}$ becomes too shallow, so the $\phi$ field does not follow the potential minimum anymore but enters the slow roll tracking regime. So the random motion of the $\phi$ field is accumulated about $\phi(a_{\text{tr}})$. If the motion of the $\phi$ driven by the potential is dominant over the stochastic effect, we can approximate that the trajectory of the $\phi$ field is solely given by the equation of motion.

Therefore, it is required that the standard deviation from the random jump is less than the $\phi_0$, and we have the following condition as
\begin{equation}
\phi_0\sim\phi(a_{\text{tr}})\sim \left(\frac{15\Lambda^2 M^5}{16 \pi^2  H^2_{\text{inf}}} \right)^{1/5} > \frac{H_{\text{inf}}}{2\pi} \sqrt{\Delta N} ,
\end{equation}
where the $\Delta N$ is the number of e-folding between the $a_{\text{tr}}$ and $a_{\text{end}}$. If we choose $\Delta N=60$ as a conservative manner, then the upper bound on $H_{\text{inf}}$ is given as 
\begin{equation}
     H_{\text{inf}} \lesssim 16\, \text{GeV}.
\end{equation}

\begin{figure}[t]
\centering
\includegraphics[width=0.9 \linewidth]{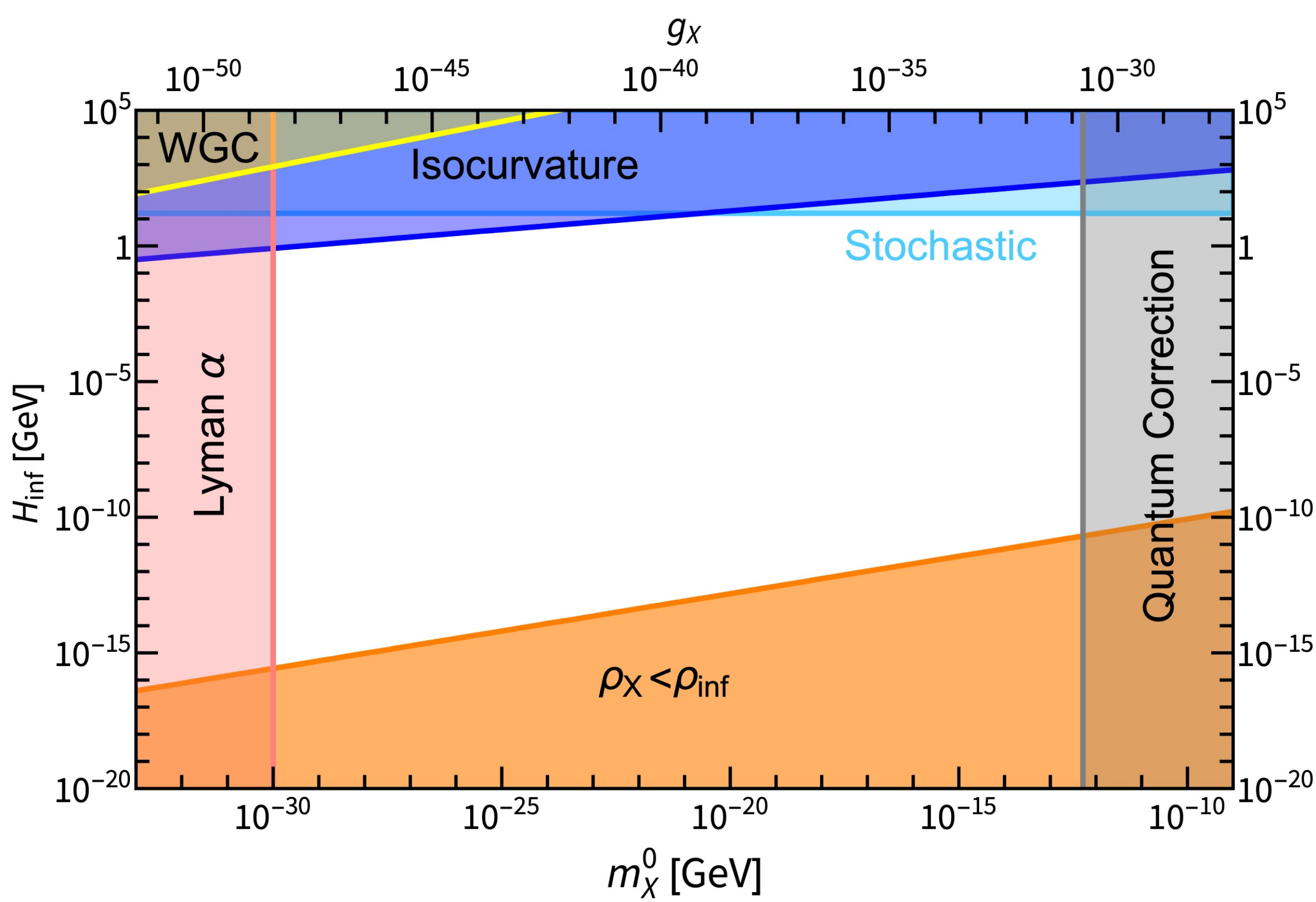}
    \caption{Constraints on the scenario with $\alpha=1$ when $\rho_X(a_0)$ is the expected dark matter density. The blue region is excluded by the isocurvature fluctuation constraint. The red region is excluded by the $\rho_X^{}< \rho_\text{inf}^{}$ (Back reaction of the $\rho_X$) during the inflation. The light blue region is constrained by the stochastic correction to the $\phi$ field. The gray region is disfavored by the quantum correction in the gauged quintessence potential \cite{Kaneta_2023}. The light red region is excluded by the Lyman-$\alpha$ forest \cite{Irsic:2017yje,Kobayashi:2017jcf}. The yellow area is disfavored by the weak gravity conjecture. The present $\phi$ value is $\phi_0 \approx 3 \times 10^{18}$ GeV.} 
    \label{fig:constraint}
\end{figure}

Figure~\ref{fig:constraint} summarizes various constraints discussed in this section and also shows other constraints from the literature. The orange region shows the constraint from the backreaction of the $X$ boson to inflation. Since the inflaton energy density is proportional to $H_{\text{inf}}$, this gives the lower bound on $H_{\text{inf}}$. On the other hand, the isocurvature fluctuation of $X$ boson and stochastic motion of the $\phi$ field is proportional to $H_{\text{inf}}$, so they give the upper bound on $H_{\text{inf}}$ as blue and light blue constraints. The gray region is excluded from the quantum correction of the $X$ boson to the $\phi$ potential \cite{Kaneta_2023}. This constraint is necessary to protect the slow roll of the $\phi$ field. However, this might be alleviated, if some unknown mechanism can cancel the gauge boson loop diagrams or one takes the Ratra-Peebles potential as a quantum effective potential which already includes all the quantum corrections. The red region on the left is the constraint from the Lyman-$\alpha$ forest. If the mass of dark matter is extremely small, the wave nature of dark matter can suppress the growth of the power spectrum below its de Broglie wavelength. (See Ref.~\cite{Hui:2021tkt} for a review.) This gives a lower bound for the dark matter component in $10^{-31}-10^{-29}$ GeV range. The ultra light boson of mass range roughly $10^{-30} - 10^{-20}$ GeV may source the black hole superradiance \cite{Arvanitaki:2009fg,Arvanitaki:2010sy,Pani:2012vp,Brito:2015oca,Baryakhtar:2017ngi,Davoudiasl:2019nlo,Choi:2020rgn}, which corresponds to the left half of the parameter region in the plot. The weak gravity conjecture states that a too small gauge coupling may not be congruous with the quantum gravity \cite{Arkani-Hamed:2006emk}. This leads to the following condition on the inflationary scenario as \cite{Heidenreich:2016aqi,Heidenreich:2017sim}
\begin{equation}
\begin{split}
H_{\text{inf}} \lesssim g_X^{1/3} M_{\text{Pl}}.
\label{eq:wgCon}
\end{split}
\end{equation}
The yellow area shows the disfavored region from the weak gravity conjecture.

We want to remind you that we do not expect the $X$ boson as a whole dark matter. The back reaction (orange), isocurvature (blue), and Lyman-$\alpha$ (light red) constraints become weaker if the $X$ boson constitutes only a fraction of a whole dark matter. In this sense, the constraints are conservative bounds for this scenario.

\section{Summary and discussion}
\label{Sec:Summary}
In this paper, we investigated the misalignment mechanism for a vector boson that couples to the quintessence dark energy field.
Because of the scale factor suppression of the vector boson energy density, the naive misalignment mechanism does not produce a sizable amount of the coherent vector boson. 
In the gauged quintessence model, the mass of the dark gauge boson is proportional to the quintessence field value. Thus, it can increase many orders of magnitude during cosmic evolution. So even if the energy density of the dark gauge boson gets exponential suppression from inflation, it is possible to produce a large fraction of the relic dark gauge boson energy density.  We showed that the relic density of the dark gauge boson could be comparable to the known CDM relic density because of the mass-increasing effect. 
The high-dimensional operators are assumed to be zero, though, as they should be suppressed to preserve the $X$ field dynamics in the early universe.

There are various scenarios of the misalignment mechanism for the vector boson production, which have additional ingredients such as kinetic coupling to the inflaton \cite{Nakayama:2019rhg,Nakayama:2020rka,Kitajima:2023fun} or non-minimal coupling to gravity \cite{Arias:2012az,Alonso-Alvarez:2019ixv}, where such new ingredients were primarily introduced for the sizable production of vector boson energy density. However, in the gauged quintessence model, the coupling between the vector boson and quintessence dark energy field was not introduced to explain the production of vector boson. Such a coupling naturally arises when we impose a gauge symmetry to the dark energy sector. In that sense, the gauged quintessence model can provide a rather natural misalignment production scenario for the vector boson.

Even though our analysis is based on a specific potential, the message that the mass-increasing effect of vector bosons can mitigate strong suppression of the vector boson energy density during inflation holds generically.

\begin{acknowledgments}
This work was supported in part by the JSPS KAKENHI (Grant No. 19H01899) and the National Research Foundation of Korea (Grant No. NRF-2021R1A2C2009718).
\end{acknowledgments}

\appendix
\section{Extra scenarios}
\label{Sec:Other scenarios}
\begin{figure}[tb]
    \begin{subfigure}{0.5\textwidth}
\centering
\includegraphics[width=0.99 \linewidth]{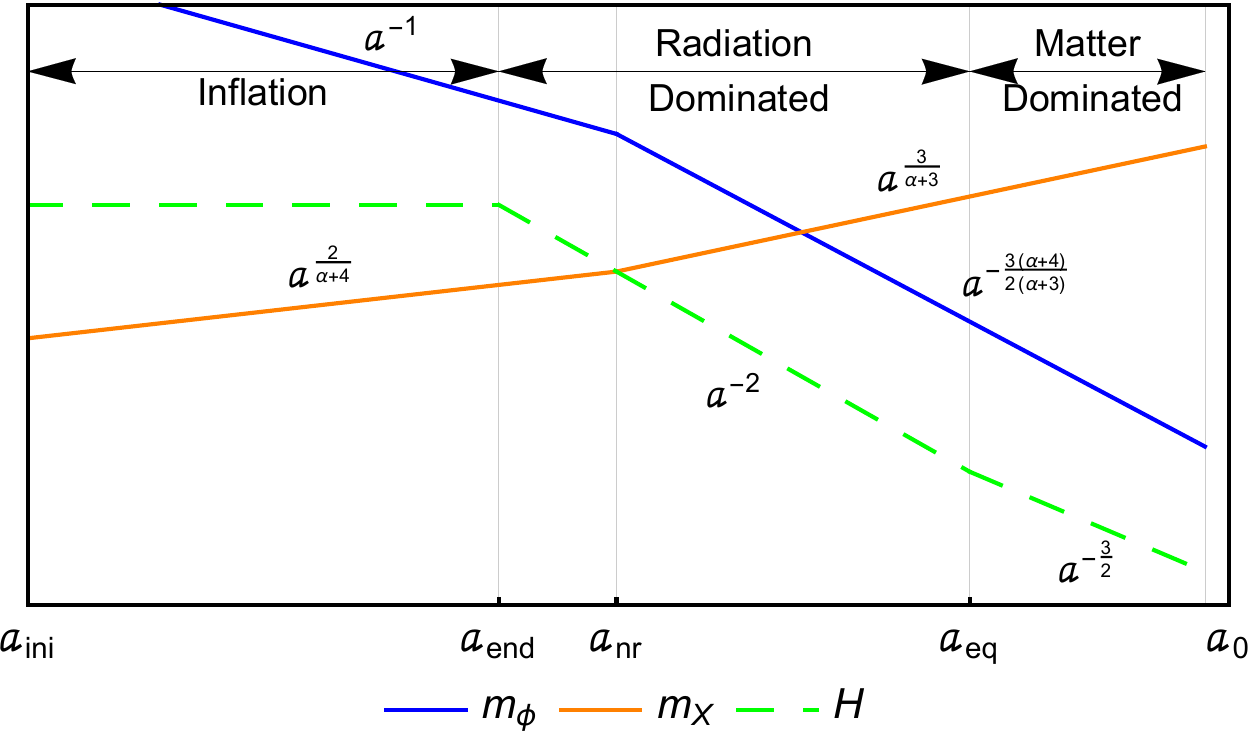}
\caption{}
\label{fig:AS1}
\end{subfigure}
\begin{subfigure}{0.5\textwidth}
\centering
\includegraphics[width=0.99 \linewidth]{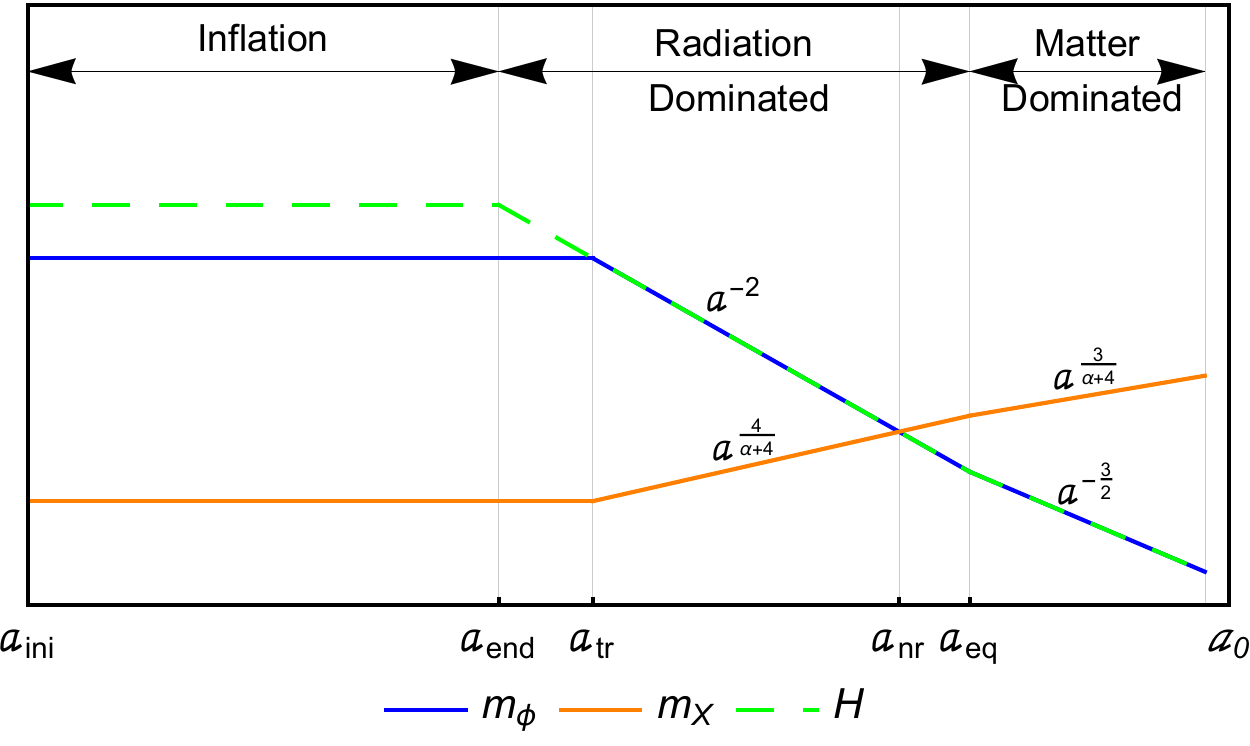}
\caption{}
\label{fig:AS2}
\end{subfigure}
\caption{Schematic description of scaling behavior of $m_\phi$, $m_X$ and $H$ for the extra scenarios in App. \ref{Sec:Other scenarios}. (a) $m_\phi \gg H$ during the entire inflationary era. (b) $m_\phi \ll H$ during the entire inflationary era. Refer to figure \ref{fig:scenario} for the explanation of the blank regime at the right of each figure.}
\label{fig:AS}
\end{figure}

In this appendix, We discuss extra scenarios which were not shown in Sec. \ref{sec:Dq}. As it was discussed, the dynamics of quintessence are determined by the relative size of $m_\phi$, $\sqrt{V_{\text{gauge}}^{\prime\prime}}$, and $H$. In the main text, we have $m_\phi \gg H$ at the beginning of inflation, and $m_\phi \sim H$ is achieved during inflation. However, other scenarios are possible such that the relations $m_\phi \gg H$ or $m_\phi \ll H$ might be maintained during the entire inflationary era\footnote{If $m_\phi \ll H$, then $\phi$ is frozen by Hubble friction. Since $H$ is constant during inflation, $m_\phi \ll H$ is maintained in the entire inflationary era, once this relationship is initially achieved.}. We describe the behaviors of $m_X$ and $m_\phi$ for each scenario in Fig. \ref{fig:AS}, and corresponding dynamics are summarized in the following texts and tables. As we discussed in Sec.~\ref{Sec:constraint}, we ignore the short dark energy dominated era with the tree-level potential domination.

\begin{itemize}
    \item Extra scenario (i): $m_\phi \gg H_\text{inf}$ during the entire inflation era. 
\end{itemize}

       \begin{table}[h!]
    \centering
    \aboverulesep=0ex
    \belowrulesep=0ex
    \begin{tabular}{|c|c|c|c|c|}
        \midrule\rule{0EM}{1.1EM}
        & $\overline X $ & $ \phi $ & {$\rho_X$}& $\rho_\phi$\\
        \midrule\rule{0EM}{1.4EM}
        $ a_\text{ini}<a<a_\text{end} $ & $\propto a^{-1}$ & $\propto a^{\frac{2}{\alpha+4}}$ &  $\propto a^{-\frac{2\alpha+4}{\alpha+4}}$ & $\propto a^{-\frac{2\alpha+4}{\alpha+4}}$ \\
        \midrule  \rule{0EM}{1.4EM}  $a_\text{end}<a<a_\text{nr}$ & $\propto a^{-1}$& $\propto a^{\frac{2}{\alpha+4}}$        & $\propto a^{-\frac{2\alpha+4}{\alpha+4}}$ & $\propto a^{-\frac{2\alpha+4}{\alpha+4}}$\\
         \midrule  \rule{0EM}{1.4EM}  $a_\text{nr}<a<a_\text{eq}$ & $\propto a^{-\frac{3(\alpha+4)}{2(\alpha+3)}}$ & $\propto a^{\frac{3}{\alpha+3}}$ & $\propto a^{-\frac{3\alpha+6}{\alpha+3}}$ & $\propto a^{-\frac{3\alpha+6}{\alpha+3}}$\\
        \midrule \rule{0EM}{1.4EM}
        $a_\text{eq}<a<a_0$  & $\propto a^{-\frac{3(\alpha+4)}{2(\alpha+3)}}$ & $\propto a^{\frac{3}{\alpha+3}}$ &   $\propto a^{-\frac{3\alpha+6}{\alpha+3}}$ & $\propto a^{-\frac{3\alpha+6}{\alpha+3}}$ \\
        \midrule
    \end{tabular}%
    \caption{Dynamics for the extra scenario (i). }
    \label{tab:sc1}
\end{table}%
In this scenario, the initial $V_{\text{gauge}}$ is the largest of all three scenarios.
Therefore, it takes a long time for $m_{\phi}$ to be comparable to $H$. Since the $\phi$ field follows minimum of the potential before $a_\text{tr}$, the scaling of the $m_{\phi}$ is given as $m_\phi \propto a^k$ with $k\ge -2$. This means that $m_\phi$ is always larger than $H$ even during the radiation-dominated epoch, so we get $a_\text{tr}>a_{\text{eq}}$. 
To obtain the accelerated expansion, $a_\text{tr}$ should be smaller than $a_0$; since the slow roll of $\phi$ cannot be achieved if the $\phi$ field runs along the potential minimum. However, an explicit calculation shows that $a_{\text{tr}}>a_0$, so the $\phi$ cannot be the dark energy. 

\begin{itemize}
\item Extra Scenario (ii): $m_\phi \ll H_\text{inf}$  during the entire inflation era. 
\end{itemize}

       \begin{table}[h!]
    \centering
    \aboverulesep=0ex
    \belowrulesep=0ex
    \begin{tabular}{|c|c|c|c|c|}
        \midrule\rule{0EM}{1.1EM}
        & $\overline X $ & $ \phi \propto m_X$ & {$\rho_X$} &$\rho_\phi$ \\
        \midrule\rule{0EM}{1.4EM}
        $ a_\text{ini}<a<a_\text{end} $ & $\propto a^{-1}$ & $\sim \text{constant}$ &  $\propto a^{-2}$ & $\sim \text{constant}$ \\
        \midrule  \rule{0EM}{1.4EM}  $a_\text{end}<a<a_\text{tr}$ & $\propto a^{-1}$& $\sim \text{constant}$        & $\propto a^{-2}$ & $\sim \text{constant}$   \\
         \midrule  \rule{0EM}{1.4EM}  $a_\text{tr}<a<a_\text{nr}$ & $\propto a^{-1}$ & $\propto a^{\frac{4}{\alpha+4}}$ & $\propto a^{-\frac{2\alpha}{\alpha+4}}$ & $\propto a^{-\frac{4\alpha+8}{\alpha+4}}$  \\
        \midrule \rule{0EM}{1.4EM}
        $a_\text{nr}<a<a_\text{eq}$ & $\propto a^{-\frac{3\alpha+16}{2\alpha+8}}$ & $\propto a^{\frac{4}{\alpha+4}}$ &  $\propto a^{-\frac{3\alpha+8}{\alpha+4}}$ & $\propto a^{-\frac{4\alpha+8}{\alpha+4}}$  \\
        \midrule  $a_\text{eq}<a<a_0$ & $\propto a^{-\frac{3\alpha+15}{2\alpha+8}}$ & $\propto a^{\frac{3}{\alpha+4}}$  & $\propto a^{-\frac{3\alpha+9}{\alpha+4}}$ & $\propto a^{-\frac{3\alpha+6}{\alpha+4}}$ \\
        \midrule
    \end{tabular}%
        \caption{Dynamics for the extra scenario (ii).}
        \label{tab:sc3}
\end{table}%

For the second extra scenario, both $m_X$ and $m_\phi$ are smaller than $H_\text{inf}$ during inflation. For this scenario to work, $H_\text{inf}$ should be large and $g_X$ should be very small. However, this scenario could not provide sufficient density of $\rho_X(a_0)$. To have a sensible scenario for $\alpha=1$, $g_X$ should be smaller than $10^{-61}$ and it is disfavored by Lyman-$\alpha$ spectrum or weak gravity conjecture.

\section{Growth of the fluctuations}\label{Sec:Growth of the fluctuations}

In this appendix, we discuss the evolution of the longitudinal mode, especially for the scenario in Sec. \ref{Sec:dynamics}. The Fourier component of the $X$ field is written as
\begin{equation}
    \mathbf{X}_{\mu}(\vec{k},\tau) =\int d^3x X_{\mu}(x)e^{-i\vec{k}\cdot \vec{x}},
\end{equation}
where $\tau$ is the conformal time defined by $d\tau = dt/a$. Then, the longitudinal part of the $X$ boson is obtained from the following decomposition, $ \vec{X} = \vec{X}_T+\hat{k}X_L$ with $\hat{k}\equiv \vec{k}/|\vec{k}|$, and $\vec{k}\cdot \vec{X}_T=0$.

Hence, the action for the longitudinal mode is written as
\begin{equation}
    S_L=\frac{1}{2}\int\frac{d\tau d^3k}{(2\pi)^3}\left(\frac{a^2m_X^2}{k^2+a^2m_X^2}|\mathbf{X}_L^{\prime}|^2 -a^2m_X^2|\mathbf{X}_L|^2 \right),
\end{equation}
where $\prime$ denotes the derivative with respect to $\tau$ unlike the main text. Note that the kinetic term is not canonical. One can obtain the canonically normalized action by using the redefined field $\mathbf{X}^c_L \equiv am_X^{}\mathbf{X}_L/\sqrt{k^2+a^2m_X^2}=g \mathbf{X_L}$,
\begin{equation}
    S_L= \frac{1}{2}\int \frac{d\tau d^3k}{(2\pi)^3} \left(|\mathbf{X}^{c\prime}_L|^2-\left(k^2+a^2m_X^2-\frac{g^{\prime\prime}}{g}\right)|\mathbf{X}_L^c|^2 \right),
\end{equation}
where
\begin{equation}
    \frac{g^{\prime\prime}}{g}= \frac{k^2}{k^2+a^2m_X^2}\left(\frac{a^{\prime\prime}}{a}+\frac{\phi^{\prime\prime}}{\phi}+\frac{2a^{\prime}\phi^{\prime}}{a\phi}-\frac{3a^2m_X^2}{k^2+m_X^2}\left(\frac{a^{\prime}}{a}+\frac{\phi^{\prime}}{\phi}\right)^2\right).
\end{equation}
 Then, the mode function $X_L^c$ is defined as
\begin{equation}
    \mathbf{X}^c_L(\vec{k},\tau) = X_L^c(\vec{k},\tau) a^{}_{\vec{k}}+ X_L^{c\ast}(\vec{k},\tau) a^{\dagger}_{-\vec{k}}, \quad [a_{\vec{k}_1}^{},a^{\dagger}_{\vec{k}_2}]= (2\pi)^3\delta^{(3)}(\vec{k}_1-\vec{k}_2).
\end{equation}
Therefore, the equation of motion for the mode function is
\begin{equation}
    X_L^{c\prime\prime} +\left(k^2+a^2m_X^2-\frac{k^2\mathcal{H}^2}{k^2+a^2m_X^2}\left((n+2)(n+1)-3\left(n+1\right)^2\frac{a^2m_X^2}{k^2+a^2m_X^2}\right)\right) X_L^c=0,
    \label{eq:modeEom}
\end{equation}
where $\mathcal{H}=aH$, and we assumed that $\phi \propto a^n$. Due to its complexity, this equation does not have an exact analytic solution. But the scales of $\mathcal{H}$, $k$, $am_X^{}$ are different in most eras, so one can solve this equation by only considering the dominant terms. 

In the scenario in the main text, the $\phi \propto a^{2/5}$ for $a<a_{\text{tr}}$, and $\phi \sim \text{constant}$ for $a>a_{\text{tr}}$. So we approximate that $n=2/5$ for $a<a_{\text{tr}}$, and $n=0$ for $a>a_{\text{tr}}$. Also, the $a_{\text{nr}}>a_{\text{end}}$ is taken for the scenario in the main text, so the $\mathcal{H}\gg a m_X^{}$ is satisfied during the inflation. We can divide possible eras as in Tab.~\ref{tab:5cases}.
\begin{table}[h]
    \centering
    \begin{tabular}{|c|c|}\hline
      $(a-1): k \gg \mathcal{H}\gg am_X^{},\, a<a_{\text{tr}}$   & $(a-2): k\gg\mathcal{H}\gg am_X^{},\, a>a_{\text{tr}}$ \\\hline
      $(b-1): \mathcal{H} \gg k \gg am_X^{},\, a<a_{\text{tr}}$   & $(b-2): \mathcal{H} \gg k \gg am_X^{},\, a>a_{\text{tr}}$ \\ \hline
        & $(c-2): \mathcal{H} \gg am_X^{}\gg k,\, a>a_{\text{tr}}$\\ \hline
    \end{tabular}
    \caption{5 possible eras for the evolution of longitudinal modes. We assume the Bunch-Davies initial condition \cite{Bunch:1978yq}, so the evolution of longitudinal mode starts from the $(a-1)$, and proceeds to the next era onto the down or right side. One can check that $(c-1): \mathcal{H} \gg am_X^{}\gg k,\, a<a_{\text{tr}}$ cannot occur in the scenario in the main text when $\alpha=1$ and $\rho_X^{}(a_0)/\rho_{\text{CDM}}(a_0) \sim 1 $.}
    \label{tab:5cases}
\end{table}

Since the observable scale of the universe should leave the horizon during inflation, the era at the end of inflation should be either $(b-2)$ or $(c-2)$. Therefore, possible cosmic histories during the inflation are 
\begin{itemize}
    \item $s_1$: $(a-1)\rightarrow (a-2) \rightarrow (b-2)$,
    \item $s_2$: $(a-1)\rightarrow (a-2) \rightarrow (b-2) \rightarrow (c-2)$,
    \item $s_3$: $(a-1)\rightarrow(b-1)\rightarrow(b-2)$,
    \item $s_4$: $(a-1)\rightarrow(b-1)\rightarrow(b-2)\rightarrow(c-2)$.
\end{itemize}

Let's first solve Eq.~\eqref{eq:modeEom} for $k\gg am_X^{}$ case. This will give the general solution for the $(a-1)$, $(a-2)$, $(b-1)$, $(b-2)$. Using that $\mathcal{H}=-1/\tau$\footnote{The inflation ends at $\tau=0$, and $\tau$ is negative during the inflation.}, we get
\begin{equation}
    X_L^{c\prime\prime}+\left(k^2+a^2m_X^2-\frac{1}{\tau^2}\left(\nu_{\chi}^2-\frac{1}{4}\right)\right)X_L^c=0,\quad \nu_{\chi}^2=\begin{dcases}
    \frac{361}{100}\quad &(a<a_{\text{tr}}),\\   \frac{9}{4}  \quad &(a>a_{\text{tr}}).
    \end{dcases}
\end{equation}
The solution is
\begin{equation}
    X_L^c=\sqrt{-\tau} \left(C_{1,\nu_\chi} H^{(1)}_{\nu_{\chi}}(-k\tau)+C_{2,\nu_\chi} H^{(2)}_{\nu_{\chi}}(-k\tau)\right),
    \label{eq:cSolutions}
\end{equation}
where the $H^{(1)}(x)$ ($H^{(2)}(x)$) is the Hankel function of the first (second) kind, and $C_{1,\nu_\chi}$, $C_{2,\nu_\chi}$ are undetermined coefficients for each $\nu_\chi$. We take the Bunch-Davies initial condition \cite{Bunch:1978yq}, so the longitudinal modes in the asymptotic past infinity is the plane wave solution as
\begin{equation}
    X_L^c \sim \frac{1}{\sqrt{2k}} e^{-ik\tau}.
\end{equation}
This fixes the nonzero coefficients $C_{1,\nu_\chi}$ for the outgoing solution and sets $C_{2,\nu_\chi}$ to be zero. So the evolution of the longitudinal modes from the $(a-1)$ to $(b-1)$ is determined. In order to determine the evolution from the $(a-1)$ to $(a-2)$ or $(b-1)$ to $(b-2)$, one should connect each solutions at the $a_{\text{tr}}$ as
\begin{equation}
    C^{}_{1,19/10} H_{19/10}^{(1)}(-k\tau)\big|_{\tau=\tau_{\text{tr}}^-}=C^{}_{1,3/2}H_{3/2}^{(1)}(-k\tau)\big|_{\tau=\tau_{\text{tr}}^{+}} ,
\end{equation}
where the LHS is the $(a-1)$ or $(b-1)$ solution, and RHS is the $(a-2)$ or $(b-2)$ solution applied at $a_{\text{tr}}$.  Then one finds the following expressions for $s_1$, and $s_3$ scenarios as
\begin{equation}
    X_L^c(a_{\text{end}})\sim\begin{dcases}
        \frac{a_{\text{end}} H_{\text{inf}}}{\sqrt{2k^{3}}}  &\text{for}\quad  s_1, \\ 
         \frac{2^{19/10}}{\sqrt{2\pi}} \Gamma\left(\frac{19}{10}\right) \frac{(a_{\text{tr}}H_{\text{inf}})^{2/5}}{k_{}^{2/5}} \frac{a_{\text{end}}H_{\text{inf}}}{\sqrt{2k^3}} \quad &\text{for}\quad s_3. 
    \end{dcases}
\end{equation}

One can calculate the solution at the $(c-2)$ by considering the evolution across the $k/a \sim m_X^{}$ \cite{Nakayama:2019rhg}.  Let's denote the conformal time at the $k/a = m_X^{}$ as a $\tau_{\ast}$. Then, Eq.~\eqref{eq:modeEom} becomes
\begin{equation}
   X_L^{c\prime\prime} +\left(k^2\left(1+\frac{\tau_{\ast}^2}{\tau^2}\right)-\frac{2\tau^2-\tau_{\ast}^2}{(\tau^2+\tau_{\ast}^2)^2}\right)X_L^c=0. 
   \label{eq:approxC21}
\end{equation}
If $k \tau_{\ast}^2 < |\tau|$, one can approximate that
\begin{equation}
   X_L^{c\prime\prime} -\frac{2\tau^2-\tau_{\ast}^2}{(\tau^2+\tau_{\ast}^2)^2}X_L^c=0. 
   \label{eq:solC21}
\end{equation}
The solution to this equation is 
\begin{equation}
    X_L^c = D_1\frac{1}{\sqrt{\tau^2+\tau^2_{\ast}}} +D_2\frac{\tau^2+3\tau\tau_{\ast}}{\sqrt{\tau^2+\tau_{\ast}^2}}.
\end{equation}
Since the solution at $(b-2)$ is proportional to $a$ ($\propto 1/\tau$), the $(b-2)$ solution has to be connected to the $D_1$ solution. Then one can determine $D_2$ by matching the $(b-2)$ and $(c-2)$ solutions at $\tau \gg \tau_{\ast}$. Now, if the $|\tau_{\ast}| \gg |\tau|$, Eq.~\eqref{eq:approxC21} can be approximated as
\begin{equation}
    X_L^{c\prime\prime}+\left(k^2\frac{\tau_{\ast}^2}{\tau^2}+\frac{1}{\tau^2}\right)X_L^c=0.
\end{equation}
The solution to this equation is
\begin{equation}
    X_L^c \sim \sqrt{-\tau}\left(a_1 J_{1/2}\left(\frac{\tau}{\tau_{\ast}}\right)+a_2 Y_{1/2}\left(\frac{\tau}{\tau_{\ast}}\right) \right) \sim E_1 \tau +E_2 ,
    \label{eq:approxC22}
\end{equation}
where the $J_{1/2}$ is the Bessel function of the first kind, and $Y_{1/2}$ is the bessel function of the second kind. The $E_2$ solution is connected to the $D_1$ solution, so one can actually use the $D_1$ solution of Eq.~\eqref{eq:solC21} to the end of inflation. An explicit calculation shows that
\begin{equation}
    |X_L(a_{\text{end}})|^2\sim\begin{dcases}
        \frac{1}{2k^{3}}\frac{H^2_{\text{inf}}k^2}{m_X^{2}(a_{\text{end}})} \quad &\text{for}\; s_1,\,s_2, \\ 
        \frac{2^{19/5}}{4\pi} \Gamma^2\left(\frac{19}{10}\right) \frac{1}{k^{3}}\frac{(a_{\text{tr}}H_{\text{inf}})^{4/5}}{k_{}^{4/5}} \frac{H^2_{\text{inf}}k^2}{m_X^{2}(a_{\text{end}})} \quad &\text{for}\;  s_3,\,s_4,
    \end{dcases}
\end{equation}
where the $X_L=X_L^c/g$, and, for all cases, the $|X_L|$ is constant on time close before the $a_{\text{end}}$. Then the power spectrum of $\overline{X}_{L}$ is 
\begin{equation}
    \mathcal{P}^{}_{\overline{X}_L}(a_{\text{end}})\sim\begin{dcases} 
    \left(\frac{kH_{\text{inf}}}{2\pi a_{\text{end}} m_X^{}(a_{\text{end}})}\right)^2 \quad &\text{for}\;  s_1,\,s_2,\\
  \frac{2^{14/5}}{\pi} \Gamma^2\left(\frac{19}{10}\right) \left(\frac{a_{\text{tr}}H_{\text{inf}}}{k}\right)^{4/5} \left(\frac{kH_{\text{inf}}}{2\pi a_{\text{end}} m_X^{}(a_{\text{end}})}\right)^2 \quad &\text{for}\;  s_3,\,s_4.
    \end{dcases}
\end{equation}
Here, the $s_1$, $s_2$ ($s_3$, $s_4$) corresponds to $a_{\text{tr}}<k/H_{\text{inf}}$ ($a_{\text{tr}}>k/H_{\text{inf}}$).

So, we found the solution of the $X_L^c$ to the end of inflation for various cases. In order to find how these mode functions evolve after the inflation, it is convenient to use the redefined field $\overline{X}_L \equiv X_L/a$, and write the equation of motion with a cosmic time. Then, one can write the equation of motion of $\overline{X}_L$ as 
\begin{multline}
    \ddot{\overline{X}}_L+\left(3H+\frac{2k^2}{k^2+a^2m_X^2}\left(H+\frac{\dot{\phi}}{\phi}\right)\right) \dot{\overline{X}}_L\\ +\left(a^2m_X^2+ 2H^2+\dot{H}+\frac{2k^2H}{k^2+a^2m_X^2}\left(H+\frac{\dot{\phi}}{\phi}\right)\right)\overline{X}_L=0.
\end{multline}
Then for the super-horizon modes before the $a_{\text{nr}}$, the equation can be written as
\begin{equation}
    \begin{dcases}
         \ddot{\overline{X}}_L +(5+2n)H\dot{\overline{X}}_L +\frac{5+4n-3w_b}{2}H^2 \overline{X}_L=0 \quad &\text{for}\; k/a \gg m_X ,\\
         \ddot{\overline{X}}_L +3H\dot{\overline{X}}_L +\left(m_X^2+\frac{1-3w_b}{2}H^2 \right)\overline{X}_L=0\quad &\text{for}\; k/a \ll m_X ,
         \label{eq:afterLmodeeom}
    \end{dcases}
\end{equation}
and the solution is given by
\begin{equation}
   \overline{X}_L = \begin{dcases} 
        F_1 a^{-1} + F_2 a^{-(5+4n-3w_b)/2}&\text{for}\;k/a \gg m_X ,\\
        F_1 a^{-1} + F_2 a^{(3w_b-1)/2} &\text{for}\;k/a \ll m_X .\\
    \end{dcases}
\end{equation}
Since the $\overline{X}_L$ at the end of inflation is proportional to the $a^{-1}$, the $F_1$ solution becomes the same as the solution for the $\overline{X}$ after the inflation. Therefore, the evolution of the longitudinal mode is the same as the homogeneous mode after the inflation.

\bibliographystyle{JHEP}
\bibliography{ref.bib}

\end{document}